# Learning to Match

Themis Mavridis
Booking.com
Amsterdam, Netherlands
themistoklis.mavridis@booking.com

Pablo Estevez
Booking.com
Amsterdam, Netherlands
pablo.estevez@booking.com

Lucas Bernardi
Booking.com
Amsterdam, Netherlands
lucas.bernardi@booking.com

## ABSTRACT

Booking.com is a virtual two-sided marketplace where guests and accommodation providers are the two distinct stakeholders. They meet to satisfy their respective and different goals. Guests want to be able to choose accommodations from a huge and diverse inventory, fast and reliably within their requirements and constraints. Accommodation providers desire to reach a reliable and large market that maximizes their revenue. Finding the best accommodation for the guests, a problem typically addressed by the recommender systems community, and finding the best audience for the accommodation providers, are key pieces of a good platform. This work describes how Booking.com extends such approach, enabling the guests themselves to find the best accommodation by helping them to discover their needs and restrictions, what the market can actually offer, reinforcing good decisions, discouraging bad ones, etc. turning the platform into a decision process advisor, as opposed to a decision maker. Booking.com implements this idea with hundreds of Machine Learned Models, all of them validated through rigorous Randomized Controlled Experiments. We further elaborate on model types, techniques, methodological issues and challenges that we have faced.

## 1 INTRODUCTION

Booking.com is the world's largest online travel agent where millions of guests find their accommodation and millions of accommodation providers list their properties including hotels, apartments, bed and breakfasts, guest houses, etc. The problem of matching supply and demand can be approached from several angles:

- It can be seen as an information retrieval problem where guests search for information about accommodation available in the platform
- It can be seen as a recommender system problem, where guests get accommodation recommendations
- It can be seen as a pure Matching problem where the goal is to match guests and accommodation providers maximizing some global market efficiency criterion
- Etc.

We believe that none of these approaches on its own is enough to give an optimal experience to both guests and accommodation providers. This work focuses on extending the Recommender Systems and Information Retrieval approaches beyond the classic idea of ranking items based on a specific relevance criterion. Our approach is motivated by several interrelated issues that are difficult to address separately:

*Complex Items*: Booking an accommodation requires users to choose several elements or aspects like destination, dates, accommodation type, number of rooms, room types, refund policies, etc. These elements define a multi-dimensional space where bookable potions are located, and since not all possible combinations exist, it is not trivial to navigate; users need help to find the best combination.

*High Stakes*: Recommending the wrong movie, or the wrong song or book, or even the wrong product has a relevant impact in the consumer experience. Nevertheless, in most cases there is a way to "undo" the selection; stop listening to the song or watching the movie, even return the dissatisfying product. But once you arrived to an accommodation that does not meet your expectations, there is not much you can do about it, generating frustration and disengagement related to the platform. We want to build a system that minimizes these scenarios and maximizes trust with the platform.

*Supply Constraint*: Accommodations have finite availability which is limited and dynamic. The interaction of it with prices directly affects guest preferences and the behavior of accommodation providers. Factoring this aspect in the shopping experience is arguably the hardest challenge.

*Infinitesimal Queries*: Guests searching for accommodations barely specify a destination city, maybe dates and number of guests. We want to build a platform that can give satisfying shopping and accommodation experiences starting from this almost zero-query scenario.

*Continuous Cold Start*: Guests are in a continuous cold start state. Most people only travel once or twice every year. By the time they come back to our web site their preferences might have changed significantly; long in the past history of users is usually irrelevant.

Furthermore, new accommodations and new accommodation types are added to the supply every day, their lack of reviews and general content, such as pictures and multilingual descriptions, make it difficult to give visibility and to advertise. We want to build a system that can give a personalized experience regardless of how often a guest interacts with Booking.com and that is able to find an audience for every property from the very beginning of them visiting Booking.com[1].

*Rich content*: Accommodations have very rich content e.g. descriptions, pictures, reviews, ratings, etc. Destinations themselves also have very rich content including visitor authored pictures, free text reviews, visitors endorsements, guides, etc. This is a very powerful advertising tool, but also very complex and difficult to consume for guests. We want to build a system that successfully exploits such rich content by presenting it to users in an accessible and relevant way.

We address some of these issues and others by constructing a large set of Machine Learned models that create concrete concepts that are applied by product development teams to create features, personalize content, optimize the user interface, etc. The rest of the paper is organized as follows: Section 2 presents an overview of the different types of Machine Learned Models we use, Section 3 discusses the modeling process, Section 4 presents a few issues we found applying AB testing to this approach, and Section 5 concludes the article.

## 2 MODEL FAMILIES

In this section we give an overview of the various types of models we have built considering whether they are supply or demand centric.

### 2.1 Demand Centric Models

The demand side of the travel market displays a very unique behavior. The majority of the guests travel infrequently throughout a year. They do not interact with our platform in a daily continuous fashion as they do with the social media applications or their media consumption streams. The users are "cold" which poses a challenge when it comes to creating a personalized and optimized experience. We present four model families that allow us to effectively address this problem.

*2.1.1 Traveller Preference Models.* Users display different levels of flexibility on different aspects, from no flexibility at all to complete indifference. We consider several trip aspects like destination, property price, property location, quality, dates, and facilities among others, and build several Machine Learned models that, in combination, construct a user preference profile assigning a flexibility level to each aspect. This preference model works as a very concrete and meaningful *semantic layer*, enabling everyone involved in product development to introduce new features, personalization, persuasion messages, etc., according to the preference profile of the user. As an example consider the Dates Flexibility model that gives a measure of how flexible a user is about the dates she/he wants to travel. If a user is considered flexible, then dates recommendations might be relevant in some situations, but if the user is not flexible at all, date recommendations can only distract and confuse the user so they are not displayed at all. Other treatments might involve re-enforcing the chosen dates with relevant information like price trends, availability, etc.

*2.1.2 Traveller Context Models.* Travellers travel as couples, as families, with a group of friends, for leisure, for business, they might visit one single city for a long stay, or several cities one after the other for shorter periods. They might go by car to a close by city, or by plane to the other side of the world, etc. All of these are examples of what we call *Traveller Context*, which is a user pre-established theme of the trip that sets quite concrete constraints and requirements. Most of these contexts are not explicitly stated in our platform, and the ones that can be specified, are usually omitted by many users. Thus, predicting, or guessing the context of the current user, as early in the shopping experience as possible is highly valuable. As the Traveller Preference Models, the Traveller Context Models give a very concrete semantic layer that enables everyone in a team to create features for specific contexts. As an example consider the Family Traveller Model, that estimates how likely is that a user is shopping for a family trip. This model is useful because many Family Travellers, forget to fill in the number of children they travel with (see Figure 1a), going through a big part of the shopping process only to find out, that the chosen property is out of availability for their children. The Family Traveller Model is used to remind the user to fill in the children information as early in the experience as possible.

*2.1.3 Item Space Navigation Models.* While our users are in the process of getting "warmer" they start providing us with implicit feedback by their various actions such as scrolling, clicking, interacting with site elements linked to specific items (like photos or reviews linked to a property, or properties linked to a search), or by neglecting elements available to them. We use this implicit feedback and construct "warm" models that guide the navigation of the users in the item space (destinations, properties, dates, etc.).

The treatment provided by these models co-exists in balance with the treatment by the "cold" models - requiring zero interaction - discussed in the subsections above. The "cold" models highlight and put focus to certain items of the space while the users explore. The "warm" models capture the behavior of our users in this journey, help them navigate by altering the ranking of the items, and choose which items to display in the various situations.

For instance, "cold" models determine how we display certain properties, what kind of elements we highlight, what aspects of the properties we attempt to bring to the attention of our guests. The "warm" models capture their behavior while interacting with them, understand the interest of our guests to each interacted property and they seed our property recommendation algorithms that aim to display other properties from our long-tail inventory that could be interesting to our customers.

*2.1.4 User Interface Optimization Models.* Font sizes, number of items in a list, background colors or images, etc., all have big impact in user behaviour as measured by top-line metrics. We use machine learning to build models that directly optimize these parameters with respect to a metric of interest. We found that it is hardly the case that one specific value is optimal across the board, so our models consider context and user information to decide the best user interface.

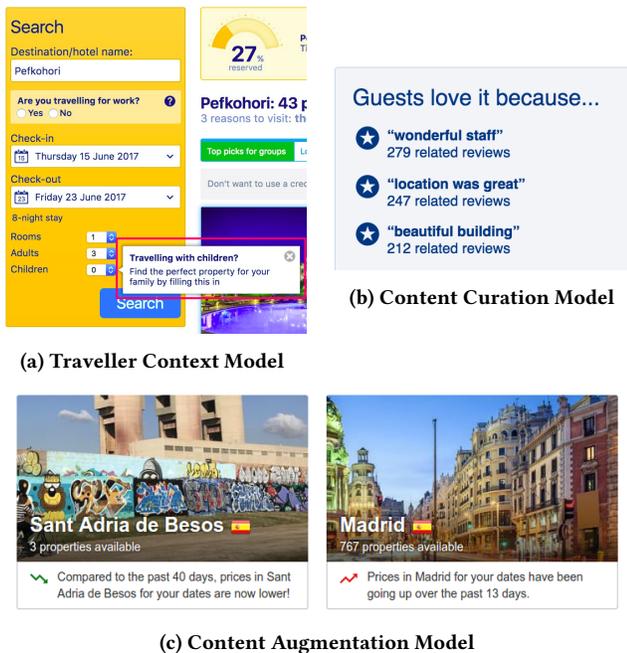

(a) Traveller Context Model

(b) Content Curation Model

(c) Content Augmentation Model

**Figure 1: Examples of Application of Machine Learning**

## 2.2 Supply Centric Models

The marketplace is long-tailed. We try to maximize the performance of the accommodations by finding those aspects of their service they excel at, and what makes them stand out from competitors. We use Machine Learning to build mainly 2 types of models:

*2.2.1 Content Curation.* We refer to content as data describing or giving information about destinations, landmarks, accommodations, special events, etc. It comes in different formats like free text, structured surveys and photos; and from different sources like accommodation managers, guests, and public data bases. It has huge potential since it can be used to attract and advertise guests to specific cities, dates or even properties. But, is also very complex, noisy and vast, making it hard to be consumed. The process of making content accessible to humans is what we refer to as *Content Curation* (Figure 1b), and Machine Learning is at the center of it. We give a few application examples of the idea of Content Curation.

- *Review Summarization*: Booking.com has collected over 125M reviews in about 1.5M properties, which contains highly valuable information about the service a particular accommodation provides and a very rich source of selling points. We use Machine Learning to "curate" reviews, constructing brief and representative summaries of the positive or outstanding aspects of an accommodation.
- *Destination Highlights*: Users on our platform not only review our properties, they also leave their opinions about what they found interesting in the destinations they travelled. When collecting that feedback, we offer both structured surveys (predefined set of tags like shopping, food, museums, etc.) and free text. Machine Learning is used to 'curate' such data into Destination Highlights that are used, among other things, to enforce a Destination selection, highlight unique aspects of a city or even give very detailed recommendations about how to make the most out of it.
- *Photo Tagging*: Booking.com collects photo content from its guests and properties, as well as external and internal sources. This content can cover anything from surroundings of a property, clear shots of its interior, details of a toilet faucet, experiences from their guests. As described in our blog[7], we use Machine Learning to label accommodation pictures with tags that indicate "what is this picture about". These tags are then used for all kind of treatments like photo gallery optimization, accommodation's facility visualization, etc.
- *Facilities Highlighting*: Each property provides our guests with multiple amenities.The list can be quite long. Machine Learning is used to highlight the facilities that provide a more clear representation of what a property offers to our guests.

*2.2.2 Content Augmentation.* The whole process of users browsing, selecting, booking, and reviewing accommodations, puts to our disposal implicit signals that allow us to construct deeper understanding of the services and the quality a particular property or destination can offer. We derive attributes of a property, destination or dates, and use them to *augment* the explicit service offer. To make this concrete, we give a few examples:

- *Value Deals*: Booking.com has a wide selection of properties, offering different levels of value in the form of amenities, location, quality of the service and facilities, policies, and many other dimensions. One property may charge more than another and still be a better deal given that it offers much more value to the user. Users need to assess how the price asked for a room relates to the value they will obtain by paying it. Value deals icons simplify this process by highlighting properties which offer a high value for the price they are asking, as compared to the options available to the user. This icons are placed after analysing the prices and value propositions in the market in which each property is placed.
- *Family Friendly Hotels*: Families are a very special segment of our users with particular requirements in terms of amenities, surroundings and other characteristics of the properties they visit. By modeling booking patterns of different traveler types we can identify which properties meet the requirement-set that would make them an ideal stay for a family, and even target it for different family types.
- *Price Trends*: Depending on the anticipation of the reservation, the specific travelling dates and the destination among other things, prices display different dynamics. Since we have access to thousands of reservations every day, we can build a very accurate model of the price trend of a city for a given time and travelling dates. When the model finds a specific trend, we inform the users to help them make a better decision which might encourage her to choose a destination and dates that look like an opportunity, or discourage particular options in favor of others. Note that in this case, the augmented item is not an accommodation but a destination plus dates (see Figure 1c).

## 3 MODELING

We found several challenges when applying Machine Learning to model the concepts mentioned in the previous section. Here, we describe a few that we consider of high importance and/or interesting as source of future research.

### 3.1 Constructing Machine Learning Problems

The Problem Construction process is a process that takes as input a business case or concept and outputs a well defined prediction problem (usually a supervised machine learning problem), such that a good solution effectively models the given business case or concept. Usually the point at which the prediction needs to be made is given (although there might be several), what fixes the features space universe, this is well known and covered by the Feature Engineering literature. But the target variable and the observation space are not always given, they need to be carefully constructed. As an example, consider the Dates Flexibility model mentioned before. It is not obvious what flexibility means: does it mean that a user is considering more alternative dates than a typical user? or that the dates she will end up booking are different to the ones she is looking at right now; or maybe it means that a visitor is willing to change dates but only for a much better deal, etc. For each of these definitions of flexibility a different learning setup can be used, for example, we could learn to predict how many different dates the user will consider applying regression to a specific dataset composed by users as observations, or to estimate the probability of changing dates by solving a classification problem, where the observations are searches, and so on. These are all constructed machine learning problems, that, when solved, output a model of the Dates Flexibility of a user.

It is not easy to formalize the Problem Construction Process, our approach is to follow simple heuristics, that consider among others, the following aspects:

(1) *User Base Impact*: this applies mainly to binary classifiers where the positive class is the class of users that will be treated (the dates flexibility model is an example). If this class proportion is very low, that means the proportion of impacted traffic by the ultimate treatment will also be very low, which in turn means commercial impact and statistical power of the experiment testing the new feature will be low.
(2) *Learning Difficulty*: when modeling these very subjective concepts, target variables are not given as ground truth, they are constructed. Therefore, some setups are harder than others from a learning perspective. Quantifying how learnable a target variable as a function of the available features is not straightforward. For regression and classification problems the Bayes Error is a good estimate since it only depends on the data set, but it is not easy to compute in all cases. Another popular approach that works well for ranking problems is to compare the performance of simple models against trivial baselines like random and popularity. Setups where simple models can do significantly better than trivial models are preferred.
(3) *Data to Concept Match*: some setups use data that is closer to the concept we want to model. For example, for the Dates Flexibility case we could create a data set asking users if they know the dates they want to travel on, and then build a model to predict the answer. This would give a very straightforward classification problem, that, compared to the other options, sits much closer to the idea of Dates Flexibility.
(4) *Selection Bias*: Constructing label and observation spaces can easily introduce selection bias. An unbiased problem would be based on observations that map 1 to 1 to predictions made when serving, but this is not always possible or optimal. Continuing the Dates Flexibility case, the regression problem based on users or sessions as observations and numbers of different dates clicked as target, has no obvious bias since every user or session is part of the observation space. On the other hand, the classification problem that estimates the probability of a user booking what she/he is looking at vs booking something else, introduces a very clear and strong bias towards bookers. Diagnosing this bias is straightforward: consider a sample of the natural observation space (users or sessions in the dates flexibility case), we can then construct a classification problem that classifies each observation into the class of the observations for which a target variable can be computed and the class of the observations for which a target variable cannot be computed, if this classification problem is easy (in the sense that a simple algorithm performs significantly better than random) then the bias is severe and must be addressed. Correcting for this type of bias is no obvious. Techniques like Inverse Propensity Weighting[6] and Doubly Robust[2] are helpful in some cases, but they offer very strong guarantees only under very strong assumptions, and they require at least one extra model to build (the propensity model). Other approaches that have been applied successfully but not systematically are methods from the PU-Learning[5] and Semi Supervised Learning fields.

In many situations many problems are good candidates; when that is the case, all of them are evaluated through AB tests.

### 3.2 Label-free Evaluation

When models are serving predictions, it is crucial to monitor the quality of predictions or estimations they make, but this poses a few challenges:

*Incomplete feedback*: In many situations labels or target variables cannot be observed, for example consider a model that predicts whether a customer will ask for a "special request", its' predictions are used while the user shops (search results page and hotel page), but we can only assign a true label to predictions that were made for users that actually booked, since it is at booking time when the special request can be filled in.

*Delayed feedback*: In other cases the actual label is only observed many days or even weeks after the prediction is made. Consider a model that predicts whether a user will submit a review or not, we might make use of this model at shopping time, but the true label will be only observed after the guest completes the stay and.

Therefore, label-dependent metrics like precision, recall, etc, are inappropriate, which leads us to the following question: what can we say about the quality of a binary classifier by just looking at the predictions it makes when serving? To answer this question we apply what we call Response Distribution Analysis which is a method composed by a set of heuristics that point to potential pathologies in the model. The method is based on the Response Distribution Chart (RDC), which is simply histogram of the output of the model. The simple observation that the RDC of an ideal model should have one peak at 0 and one peak at 1 (with heights given by the class proportion) allows us to characterize typical patterns that signal potential issues in the model, a few examples are:

(1) A smooth unimodal distribution with a central mode might indicate high bias in the model or high Bayes error in the data
(2) An extreme, high frequency mode might indicate defects in the feature layer like wrong scaling or false outliers in the training data
(3) Non-smooth, very noisy distributions point to too sparse models
(4) Smooth bimodal distributions with one clear stable point are signs of a model that successfully distinguishes two classes

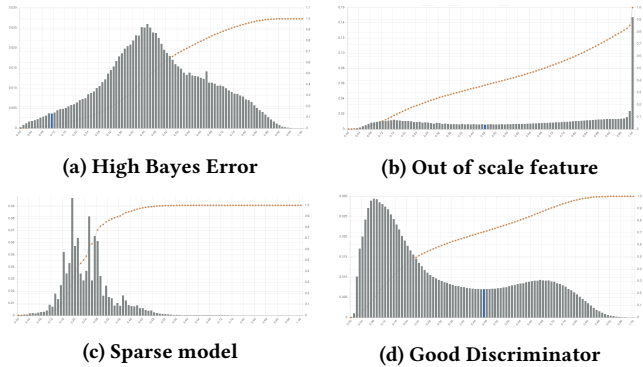

(a) High Bayes Error  (b) Out of scale feature

(c) Sparse model  (d) Good Discriminator

Figure 2: Examples of Response Distribution Charts

Figure 2 illustrate these heuristics. The rationale behind these heuristic is that if a model cannot assign different scores to different classes then it is most likely failing at discriminating one from another, small changes in the score should not change the predicted class. It is not important where the stable point is (which could indicate calibration problems), it only matters that there is one, since the goal is to clearly separate two classes, one that will receive a treatment one that will not. Without going in full detail, these are the advantages this method offers:

(1) it can be applied to any scoring classifier
(2) it is robust to class distribution. In extreme cases, the logarithm of the frequency in the RDC is used to make the cues more obvious
(3) it addresses the Incomplete Feedback issue providing Global Feedback since the RDC is computed considering all predictions
(4) it addresses the Delayed Feedback issue providing Immediate Feedback, since the RDC can be constructed as soon as a few predictions are made

(5) It is sensitive to both class distribution and feature space changes, since it requires very few data points to be constructed
(6) it can be used for multi-class classification when the number of classes is small by just constructing one binary classifier per class that discriminates between one class and the others (one vs all or one vs rest)
(7) It offers as a label-free criterion to choose a threshold for turning a score into a binary output. The criterion is to simply use any point in between the 2 modes, if the region is large, then one can choose to maximize recall or precision using the lower and upper bound of that region respectively. This is very useful when the same model is used in various points of the system like hotel page or search results page, since they have different populations with different class distribution.

The main drawbacks are:

(1) It is a heuristic method, it cannot prove or disprove a model has high quality
(2) It does not work for estimators or rankers

In practice, Response Distribution Analysis has proven to be a very useful tool that allows us to detect defects in the models very early.

### 3.3 Offline Metrics vs Business Value

It often happens that new models or algorithms, improve with respect to a baseline successful model in terms of off-line, label-dependent metrics like precision, recall, RMSE, etc. but fail to produce business impact. There are many reasons that cause this effect, we highlight the most important ones:

Selection Bias: As discussed before, many models are trained only using a subsample of the target population for which ground truth data is available. Therefore, when computing classic machine learning metrics, the metrics themselves are biased, turning them in bad predictors of the performance of the model once in production.

Correlated Models: A new model might improve the performance on a task with respect to an existing baseline. But the incremental impact is given by the fraction of cases where the models disagree. As the models get better, this proportion decreases, to extremes where the effects are very difficult to detect in an A/B test. (See section 4.1)

Loose Model-Treatment Coupling: Many models play a key role in a website functionality, but how they are actually used is what drives most of the business impact. Model usage ranges from minimal things like personalizing a color, to more prominent things like what type of accommodations or destinations the users see, or even what functionality the users see. In many cases a good-enough model is sufficient to unlock the value of a specific functionality, we call these models, loosely coupled models denoting the fact that the relationship between their accuracy an the commercial value is inferior: only big improvements in accuracy bring visible commercial impact. When working with loosely coupled models, the correlation of accuracy and commercial impact is low, discouraging model iteration.

## 4 EXPERIMENTATION

Experimentation and Randomized Controlled Trials (RCTs) are ingrained into Booking.com culture. We have built our own experimentation platform which democratizes experimentation to make sure that anything we do and can be put to test through a RCT is indeed put to test to validate its hypothesis and assess its impact [3]. Machine learning models also benefit from this process. The in-house infrastructure allows us to select, tune and learn from our models through experiments. Out of the standard experimentation practices, machine learning models present special challenges, opportunities and approaches which we will touch upon in the following paragraphs.

### 4.1 Correlated Models

When two models are compared using A/B testing, the users are part of the experiment only if the the models disagree as much as to give a different treatment; when that is the case, a coin is flipped to decide if the user is assigned to treatment or control group. But if both models agree on whether the user should be treated or not, then there is nothing to compare, and therefore the user is not part of the experiment. This implies that the number of users in the experiment grows with the probability of disagreement between the compared models. Consider a balanced binary classification problem and consider model $x$, a successful solution with 80% accuracy. Model $y$, solves the same problem with 90% accuracy. Only 20% of the predictions are wrong in model $x$, if all of them are corrected by model $y$, then 10% are correct in $x$ and wrong in $y$, which gives a total of 30% different predictions, that is, the probability of disagreement is 30%. This toy example illustrates the high correlation between models that improve on one another, note that this is an extremely optimistic case where all the wrong predictions of a model are corrected by the new one, in practice the correlation is much higher. Figure 4 illustrates this process and how the impact decreases as the model is improved. Some factors alleviate the effect of this phenomenon:

*Larger output spaces* give higher chances of disagreement between models. Multiclass classification problems with a treatment for each class are less vulnerable. As an extreme ranking problems are almost safe, except for the fact that the head of the ranking is usually more stable than the rest.

*Observations space*: the unit of experimentation is always the user for the demand side. But the observation space for the prediction model might be of finer grain. For example a model might predict for each search a user does, which gives again more chances of disagreement between models.

*Per-prediction impact*: if a model is used for many different tasks, then each prediction has more impact than a model that is used for a single feature.

These considerations have strong impact in the Problem Construction process, driving it towards complex output spaces, fine grain observation spaces, and high level abstractions that can be used in many different tasks.

### 4.2 Controlling Performance impact

Running machine learning models introduces additional computations to a page load, resulting in a slowdown that could negatively affect the user experience. This problem can be even larger during experiments, since additional computations may be needed in order to track model performance and to achieve fine tracking.

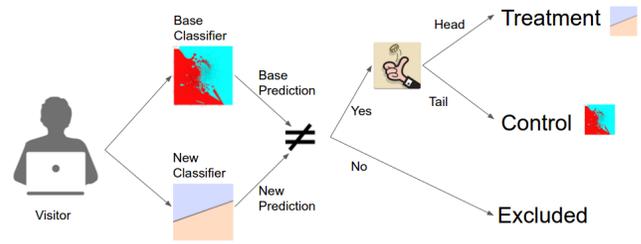

Figure 3: Testing two classifiers

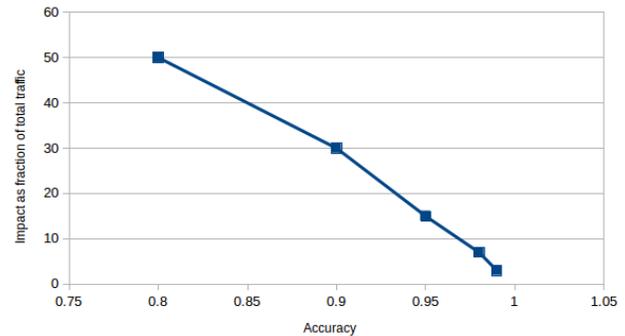

Figure 4: Upper bound of the impacted traffic for a balanced classification problem as a function of accuracy

For instance, let's consider the situation described in the previous section. We would like to compare when the prediction of two models is different and this requires essentially two separate computations. In this situation, the metrics in an experiment comparing base to a variant affected by the model will include a mixture of the effect of the element and model on the user experience (for instance, new recommendations) with the effect of the slow down on the page. In order to disentangle these two effects and obtain independent measurements of each one, we use blocked experimentation. For instance, we create an experiment with 3 variants (base, V1 and V2), expose the computation of the new model to two variants (V1, V2) and expose the actual change in V2 only. This setup enables to measure both the effect of the performance and the actual change on the business.More details on this topic can be found in [4].

### 4.3 Variability of base

Booking.com site changes constantly. We are running thousands of experiments concurrently which means that no two users see the same site. A few of those experiments become site features daily which means that over time our "base" site evolves very quickly. And on top of that, machine learning models make potentially each page load different by changing the existence, location and content of site elements in reaction to user features and interactions with the site.

On one hand this poses a complex environment where developers and testers have to assess a feature in every possible scenario. Moreover, since ML models are not always white boxes, it is hard for testers to present themselves in a way that would trick the model into providing the scenario they want to test. One simple approach we take to solve this issue is to offer the possibility to overwrite both the features sent to the model and the output returned by it, which is simplified by the fact that many models are provided as a service.

The second issue is that models must be able to react to this changing environment. Many of the model features and labels come from interactions of the users with the site, which can be affected by the continuous changes happening to it. For instance, the search parameters (number of adults, children, rooms, travel purpose, filters, sorters, etc) are important features of many models, but we continuously experiment with how we request those parameters (drop down menus, autocompletes, pre-filling, etc), the messaging around it, the options available and their order, or even the existence of a specific field.

Since we have set a model serving service, we partially tackle this problem by providing centralized monitoring for all models, in which we can observe health metrics as those mentioned in 3.2. The business impact is periodically reassessed since all areas of the site keep on being experimented with updated models, new elements and blackouts.

### 4.4 Experiments as data sources

Experiments are also a great tool for data gathering since they provide well distributed samples over our users, allow for fine grained tracking of specific segments, and monitor hundreds of metrics over each user while allowing for the creation of new ones when needed. Both existent experiments and ad-hoc ones can be used to assess the response of users to specific changes on the interface and parameters of the model used. After an experiment has run, we can look into our logs for the interaction between the user features and the parameters of the algorithms and the interface, and model their effect on any of the metrics monitored. This results in systems able to predict the optimal settings of an element in order to maximize the probability of the user exhibiting a certain behaviour, which can be in turn tested in another experiment.

## 5 CONCLUSION

In this paper, we presented our approach in how we apply Machine Learning in the marketplace of Booking.com. We described the issues that we are trying to address and the various models that we have created on both demand and supply. We analyzed important challenges on the modelling process and performing experimentation at scale. All these have lead the creation of a system characterized by the following:

*Supply Navigation* - The system helps the users to navigate the large offer of various alternatives across dimensions like where to go, when to go, how to book, etc. by making personalized recommendations at sensible points in their journey.

*Guest Preferences Discovery and Fit* - Users might not know what they want or need with respect to specific aspects of their trip. The system helps them to discover or shape their own preferences. It enables them to find items that fit their constraints and restrictions. This is achieved by displaying relevant item content or website features on specific user or market contexts, without explicitly asking for preferences.

*Decision Making Support* - Users need to make a decision at some point committing to a specific alternative. The system supports this process by highlighting persuasive attributes of an item or context in some situations or encouraging to continue shopping in others.

*Adaptive User Interface* - Booking.com provides many features in the website and mobile apps that support users with different tasks like sorting, filtering, refining, bookmarking, etc. Many aspects of the user interface are dynamic, adapting to user context and preferences, and content availability or relevance.

The main learning out of this process is that although data brings a huge opportunity to improve our platform there is a large gap between data and user experience. Machine Learning is one of the many tools that our teams use to bridge this gap.